# Prefix-based Labeling Annotation for Effective XML Fragmentation

Kok-Leong Koong, Su-Cheng Haw, Lay-Ki Soon, and Samini Subramaniam

Faculty of Computing and Informatics, Multimedia University, 63100 Cyberjaya, Malaysia

## ABSTRACT

*XML is gradually employed as a standard of data exchange in web environment since its inception in the 90s until present. It serves as a data exchange between systems and other applications. Meanwhile the data volume has grown substantially in the web and thus effective methods of storing and retrieving these data is essential. One recommended way is physically or virtually fragments the large chunk of data and distributes the fragments into different nodes. Fragmentation design of XML document contains of two parts: fragmentation operation and fragmentation method. The three fragmentation operations are Horizontal, Vertical and Hybrid. It determines how the XML should be fragmented. This paper aims to give an overview on the fragmentation design consideration and subsequently, propose a fragmentation technique using number addressing.*

## 1. INTRODUCTION

There are reasons why XML has grown to become the standard of data exchange. First of all, it is stored in a plain text format and can be easily processed by any applications and systems. It also contains features of semi-structured, self describing and human-readable document format. XML and HTML are both subset of Standard Generalized Markup Language (SGML) [1]. And, HTML is standard language that used in develop web pages and web applications since 1990s until present. Thus, it makes XML a good choice for data exchange in web environment. Consequently, XML has started to become a standard of data exchange between applications and systems. However, as the nature of XML, it is also common used in standalone applications to store metadata or application data.

Big data is a general term describe gigantic volume of data that either structured or unstructured and cannot be easily handle by standard data processing software. The emerging of smart phone, tablet and wear market has generated big volume of data and it grown exponentially in every minute or even second. The cohesiveness between these data may be low or may be highly related to each other. Thus, XML is a good choice to be used to handle these data. However, large volume data will be only effective to be stored and retrieved in distributed model as it can be making used of the parallelism processing.

What are the advantages of employ distributed design on large database? First of all, a distributed system do not need a single high-end computer system rather multiple normal specification computer system will be adequate to handle the jobs. Subsequently, it will lower the cost but still sustain the high performance on the distributed database. Secondly, the scalability of distributed system will not restrict the scale of the database to expand beyond a single computer system. Thirdly, it will increase the availability as distributed design database will be commonly replicated. This will make the database more resistance to the failure of a single computer system [2]. Lastly, it will increase the performance of the database system as it used parallelism processing to store and retrieve data from the database system [3].

There are three stages of process in distributed design of database: fragmentation, allocation and replication [4]. The focus on this paper is mainly on fragmentation. Fragmentation is a process of divide database into smaller fragments. Fragmentation contains two steps: determine a fragmentation model to be used and select a method or an algorithm to use for the fragmentation. In the first step, it determines what structure or model of fragmentation to be used. It can be horizontal, vertical or mixed. In the second step, it determines how the data should be fragmented into fragments. It also sometimes refers to fragmentation method or technique.

The rest of the paper is organized as follows. Section 2 gives the background of fragmentation model in traditional databases and also discusses on the fragmentation model in XML databases. In addition, we review the various fragmentation methods in XML databases. Section 3 presents our proposed prefix-based labellng fragmentation method. Section 4 presents our discussion. Finally, Section 5 concludes the paper.

## 2. RELATED WORKS

### 2.1. Fragmentation Model for Traditional Databases

Fragmentation on database started on traditional database such as relational database and then extended into object-oriented database. The three common models are horizontal, vertical and mixed [5].

In the relational database, data are stored in tables. Each table consists of rows and columns. Records are stored in the table in the row sequence and columns represent the data field in each row. For example, let us assume Book relation has the following columns: , *bookid*, *title, ISBN, publisher, year, price*. Horizontal fragmentation is referring a fragmentation database at the record, row or tuple level [3, 6]. As such, the resulting fragments will have the same columns *(bookid, title, ISBN, publisher, year, price)*. The number of resulted fragments is depends on the criteria set; it could be fragment based on *publisher*, or fragment based on *year* or even a range of *price*.

On the other hand, vertical fragmentation referring a fragmentation database by grouping fields or attributes of records. Each fragment contains particular field(s) of the records. Using the same example, it will split this database by grouping fields such that it might group *bookid*, *title, ISBN* and *price* fields and store in first node, while *bookid, publisher* and *year* into the second node. Both resulted fragments will have the attribute *bookid* as the primary key. This is an effective "divide and conquer" tactic that will reduce the transverse or search time on smaller fragments. For instance, we would like to search the books which have price lower than a particular amount, we only search on fragment that contain only four fields rather than six fields.

Finally, the mixed or hybrid is a combination of both horizontal and vertical fragmentation. It can be split horizontally then vertically or vice versa. The sequence of hybrid may achieve different purposes. Using the same relational example, a mixed can first split horizontally by grouping records that belong to particular level of *price*. Then, split further on the current records by splitting *bookid , title, ISBN* and *price* on other node and the rest of the data *bookid , publisher* and *year* on other node.

Object-oriented database on the other hand is different from relational database. This type of database is design more naturally fit into our real world environment. The data is stored in object form and can be illustrated in a hierarchical or tree format. Fragmentation in object oriented has increased complexity of its hierarchical structure, methods or properties within an object [7]. In term of structure, XML is quite similar to object-oriented database. Figure 1 and Figure 2 show the horizontal and vertical fragmentation on object-oriented database.

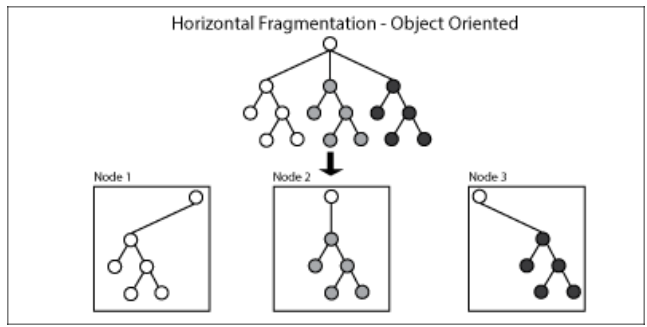

Figure 1: Horizontal Fragmentation for object-oriented database.

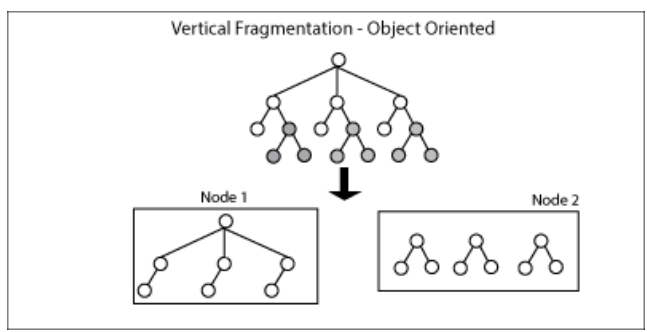

Figure 2: Vertical Fragmentation for object-oriented database.

## 2.2. Fragmentation Model for XML Databases

Generally, the fragmentation models for XML Databases can be broadly classified into Horizontal, Vertical and Hybrid [6].

Horizontal fragmentation can be achieved by selection. Selection is based on the pre defined conditions on splitting the fragments. A horizontal fragment $f_i$ is determined by the selector operator $\sigma$ of predicates p over collection of elements E in a homogeneous XML document. It can be written so that $f_i = E(\sigma_{pi})$. Assume we have a XML document constructed according to the relational database stated in the previous section. If the simple selection predicate of p1 such that */books/book/price* to be income level less than or equal 200 and p2 to be income level more than or 200, thus fragments will be written as $f_1 = E(\sigma_{p1})$ and as $f_2 = E(\sigma_{p2})$.

From Figure 3, book element with the *title Computing essentials: making IT work for you* will be then split and stored as a new XML document in node 1 as first fragment and the rest of the elements of *book* will stored in node 2 as a new XML document. After the operation, node 1 and node 2 may have DTD like <!DOCTYPE books (book*)> and <!ELEMENT book (*title, ISBN, authors, publisher, year, category, price, TableOfContent*)>. Horizontal fragmentation is recommended when the query criteria is based on particular attribute that used as selection predicate to fragmenting the XML database. In this scenario, horizontal fragmentation may reduce the transportation cost and processing time as the data is determined in a specific distributed note. Moreover, horizontal fragmentation can easily transport data between sites to improve system performance [8].

Vertical fragmentation, on the other hand, can be achieved by projection. It will split the data structure into smaller parts as particular selected child elements will be split and stored as fragment in other node. A vertical fragment $f_i$ is determined by the projection operator $\pi$ by path selection $\rho$ over collection of element E in a homogeneous XML document. It can be written so that $f_i = E(\pi_{pi})$. If the path selection $\rho i$ is *books/book/TableOfContent*, all the children elements under this tree path will be split and stored in other node. In this case, fragment $f1 = E(\pi_{p1})$ represents all contact elements in the XML document will be split and stored in node 2. And, the remaining elements will be stored in node 1.

After the operation, node 1 may have an DTD like <!DOCTYPE *books (book\*)*> and <!ELEMENT book (*title, ISBN, authors, publisher, year, category, price*)>. And, node 2 may have an DTD like <!DOCTYPE TableOfContent (Chapter\*)> and <!ELEMENT Chapter(Number, Topic)>. In order to create reference link between these two nodes, at least one reference attribute is required for the element that will able to refer back to elements that resided in other node or site [9]. Vertical fragmentation is a kind of affinity-based fragmentation. As opposed to horizontal fragmentation, this type of fragmentation does not encourage transportation of data from node to node which will trade off flexibility to affinity [8].

Hybrid fragmentation or sometimes also referring to mixed fragmentation uses both horizontal and vertical fragmentation by taken advantages of both models. A hybrid fragment $f_i$ is determined by the horizontal and vertical fragmentation implemented. It is depend on how you would like to implement the hybrid into the XML document. It can be split horizontally then vertically or vice versa. Assume you would like to do it horizontally then vertically. First fragment the document horizontally and called this fragment $f_a$. Thus, $f_a = E(\sigma_{pi})$ and from these $f_a$ fragments we further fragmented them vertically such that the hybrid fragment $f_i = f_a(\pi_{pi})$.

Assume we use *price* level as the selection condition in horizontal fragmentation, and vertically fragment further with the path *books/book/TableOfContent* as previous example. There will be 4 hybrid fragments generated for 4 nodes. After the operation, node 1 and node 2 may have DTD like <!DOCTYPE *books (book\*)*> and <!ELEMENT *book (title, ISBN, authors, publisher, year, category, price, TableOfContent)*>. Node 3 and node 4 may have DTD like <!DOCTYPE TableOfContent (*Chapter\**)> and <!ELEMENT *Chapter(Number, Topic)*>. It will look exactly like in vertical fragmentation as it final operation is using vertical fragmentation. However, each node contains only two records instead of four records using vertical fragmentation.

```xml
<?xml version="1.0" encoding="utf-8"?>
<books>
    <book>
        <title>Computing essentials: making IT work for you. compete 2012</title>
        <isbn>978-0-07-122107-8</isbn>
        <authors>
            <author>O'Leary, Timothy J</author>
            <author>O'Leary, Linda I</author>
        </authors>
        <publisher>McGraw Hill</publisher>
        <year>2012</year>
        <category>Computing</category>
        <price>98</price>
        <TableOfContent>
            <Chapter>
                <Number>1</Number>
                <Topic>Information Technology, the Internet and You</Topic>
            </Chapter>
            <Chapter>
                <Number>2</Number>
                <Topic>The Internet, the Web and Electronic Commerce</Topic>
            </Chapter>
            <Chapter>
                <Number>15</Number>
                <Topic>Your Future and Information Technology</Topic>
            </Chapter>
        </TableOfContent>
    </book>
    <book>
        <title>Murach's Android programming: training and reference</title>
        <isbn>978-1-890774-71-4</isbn>
        <authors>
            <author>Murach, Joel</author>
        </authors>
        <publisher>Mike Murach and Associates</publisher>
        <year>2013</year>
        <category>Programming</category>
        <price>186</price>
        <TableOfContent>
            <Chapter>
                <Number>1</Number>
                <Topic>An introduction to Android</Topic>
            </Chapter>
            <Chapter>
                <Number>2</Number>
                <Topic>How to use Eclipse for Android development</Topic>
            </Chapter>
            <Chapter>
                <Number>18</Number>
                <Topic>How to work with locations and maps</Topic>
            </Chapter>
        </TableOfContent>
    </book>
    <book>
        <title>Core Java. volume 1: fundamentals</title>
        <isbn>978-0-13-708189-9</isbn>
        <authors>
            <author>Wrightson, TylerHorstmann, Cay S.</author>
            <author>Cornell, Gary</author>
        </authors>
        <publisher>Prentice Hall</publisher>
        <year>2013</year>
        <category>Programming</category>
        <price>225</price>
        <TableOfContent>
            <Chapter>
                <Number>1</Number>
                <Topic>An introduction to Java</Topic>
            </Chapter>
            <Chapter>
                <Number>2</Number>
                <Topic>The Java Programming Environment</Topic>
            </Chapter>
            <Chapter>
                <Number>14</Number>
                <Topic>Multithreading</Topic>
            </Chapter>
        </TableOfContent>
    </book>
</books>
```

Figure 3: Example of XML document

## 2.3. Fragmentation Methods for XML Databases

In distributed design, fragmentation model will only define the fragmentation structure that does not determine how the data should be fragmented. A fragmentation can be done by arbitrary cutting document in to fragments horizontally, vertically, or hybrid. However, this type of fragmentation may not provide effectiveness on the query performance or storing data. Thus, some fragmentation methods have been used to achieve higher level of effectiveness. Each of this proposed method consist advantages and disadvantages against difference scenario. Basically, these methods can be categorized into four categories: structure and size, query and cost, predicate and holes and fillers (for streamed data).

## 2.3.1. Structure and Size

In this fragmentation method, document can be fragmented based on structure and size of the document. The information about the size of the document can be only obtained by transverse throughout the document. However, the structural information of the document can either be obtained from the document schemata (DTD or XML Schema) or by transverse the XML document. The main advantage of this fragmentation method is balanced the load of site or nodes processing power. At the same time, it will also reduce the transverse and search time when query the document and finally improved the effectiveness on the performance.

The disadvantage of this type of fragmentation may lead to a problem called skewed query processing problem. It is a well known problem in distributed design that indicated an unbalance on loading on particular distributed node against other nodes. This problem will slow down or overload the processing power on particular node when most of the queries are very dependent on this node. To resolve this particular skewed query processing problem, the fragments should be properly distributed according to the structure and size of the document heuristically.

The proper process of fragment document using structure and size method is to first parse the document. In other words, map the XML document into a tree structure. The parser method can be either tree-based or event-based. A tree-based parser may consume memory resources as it transverse the whole document and save all the relationship and node of these nodes in the memory and may not suitable for large document. Document Object Model (DOM) is a tree-based parser. On the other hand, event-based parser will consume less memory as it does not construct a large tree in the memory rather it will only scan particular element, attribute, content sequence in an XML document [10].

In structure and size method, event-based parser is commonly used to construct vertex/node list, structural information. This information is obtained and document can be fragment accordingly. Fragmentation method by size can be achieved by first determine the threshold size of fragment. Then, transverse throughout the XML document by determine the size of a single level child node horizontally. During the transverse, if the size of the child node including its descendants is smaller than the threshold size then continue on the next sibling child and so on until reaching the threshold size. These child nodes then will be generated as a fragment and store in a distributed node or site as illustrated in Figure 6. This scenario is vulnerable to skewed query process problem if particular fragment loading is much higher than other fragments.

Angela et al. proposed a simple top-down heuristics fragmentation method called SimpleX [11]. SimpleX method required three criteria to determine before the fragmentation. These criteria are tree-width constraint, tree-depth constraint and tree-size constraint. These criteria will determine the size of fragment. Fragment is determined when transverse down from the root element to the leaf elements (top-down). Fragment will be created upon sub tree size that fulfils the tree-size, tree-width and tree-depth constraints. Then, structure histogram is constructed to evaluate how efficient is the fragmentation generated.

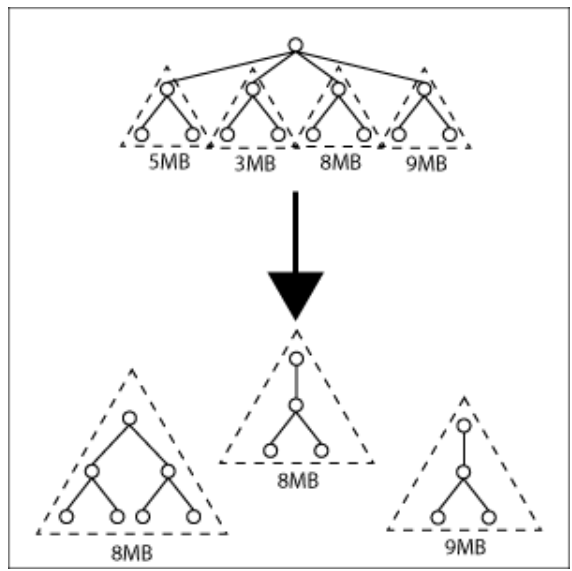

Figure 4: Fragmented by size

### 2.3.2. Query and Cost

In this method, XML document will be fragmented based on XML queries. The most common criteria that used in this fragmentation are query frequency and the cost of query.

Leykun et al. proposed vertical fragmentation model based on queries. They proposed measuring the Element Usage Matrix (EUM) and Element Affinity Matrix (EAM) [12], where by an analysis should be done on the total data access in the distributed system to determine the most frequently used queries and its frequencies. Then, the EUM will be constructed based on the elements access and queries. Subsequently, the EAM will be constructed based on the relationship between elements against the queries requested. In addition, they also proposed two heuristic modules, i,e, the Grouping Heuristic Module to group elements and Splitting Heuristic Module to decide the fragment point for the fragmentation.

Ma et al. proposed a heuristic method to fragment the XML document in horizontal fragmentation model based on the following four steps. Firstly, a horizontal fragmentation is constructed based on simple selection predicate. Then, a query tree (also refer to as query plan) is build. From the query tree, the total execution query costs (summation of storage costs and transportation costs) will be determined. Storage cost is a measure of time in retrieving data from secondary storage, while transportation cost is a measure of time to transverse time on XML document at different sites. Lastly, the minimum total query cost will determine how the document should be fragmented [13, 14]

### 2.3.3. Predicates

Predicate is commonly used in XPath. It is specific value that able to determine a specific node or a node in the document. Predicates are also common in horizontal fragmentation model. There are two types of predicates: simple selection predicates and normal selection predicates.

The simple selection predicate takes the form of *path θ v*. θ is the comparison operator which belong to the subset of $\{<. >, =, \neq, \leq, \geq,..\}$. *path* is the path expression in XML and *v* is the value [15].

Predicate in relational database is differed from XML. In relational database, predicate indicate value of the fields. However, predicate in XML is indicated by path expression similar to XPath. In the previous example, predicate in relational database can be stated as price >= 100. However, in XML, it then express in the form of /books/book/price >= 100.

### 2.3.4. Holes and Fillers

Ad-hoc fragmentation is fragmentation model for stream data. Holes and Fillers is a fragmentation method uses in Ad-hoc fragmentation. Most of the Ad-hoc fragmentation does not required document schema for document fragmentation. Fragment in this model is fragmented and mark with special identifier for reconstruction later [4]. The concept is similar to the TCP/IP fragmentation datagram in Internet layer. In TCP/IP, data is fragmented into smaller size and it will mark with sequence number in the header and it will be than able to fit into the restricted size of the packet. These fragments do not send through single path. It travels through different paths to avoid broken link on particular line. Then, it will resemble back to data.

XFrag is the framework used in holes and fillers fragmentation method. In this method, the original document is break into smaller part (fillers) and one or more holes resided in filler with special tag and contains ID to corresponding filler.

## 3. PROPOSED APPROACH: LABELING-BASED FRAGMENTATION

### 3.1 Background and Observations

Labeling scheme have been proposed to provide quick determination of structural relationships such as Parent-Child, Ancestor-Descendant and siblings (predecessor and successor) between any two nodes of XML data modeled as a tree. It has been actively exploited in the context of structural joins for centralized XML query processing. Taking the advantages of this, we observed that we could adopt the prefix-based labeling scheme to annotate the nodes to quickly identify the relationship between the fragments. For example, in vertical fragmentation of an XML tree, each of the resulting fragments is a subtree of the original XML tree. Thus, the relationship among the fragments could also be represented as a tree where a node is an XML fragment.

Labeling-based fragmentation is a new method introduced to accelerate the fragmentation of native XML. Labeling-based fragmentation can be used with other fragmentation methods such as structural and size, predicate or query and cost as introduced earlier in Section 2.3. The labeling-based fragmentation is designed by referencing label scheme in XML query and in compacting XML structure [1, 16, and 17].

Our proposed labeling-based fragmentation contains two main processes: Addressing and Fragmentation. Addressing is a process of generate a label that contains two parts of information, which are important reference in the fragmentation. The first part of the information is quite similar to the label in the label scheme which consists of information about the hierarchical structure of the data and the second part contains tag type that used in the document.

## 3.2. Addressing Process

In Addressing process, a full transverse operation is required to be executed and an *address* attribute is inserted into each element. This *address* attribute is separated by dot to indicate the level of hierarchical structure in the document and is quite similar to the label in the label scheme but it extended to contain type of tag separated by a slash (*/*). An *address* attribute in root element will be "/0" contains no first part value. In the first part of the value, each value indicate the hierarchical structure of the document starting from 1 but exclude the root element. Each level separated by dot.

The first child element within the root element will contain "1/1" that indicate this is a first record with first tag type. The second child element as sibling to first child element within the root element will contain "2/1" that indicate this is a second records with possible a first tag type. And, the consecutive first child element within the first child element will contain "1.1/2" that indicate this is the first record of first child element with second tag type.

Using Figure 3 as an example, the *<books>* element contain address value of "/0". The first child of *books*, *book*, contain value of "1/1" with tag type 1, which indicates *book* tag. Consequently, the next element, *title* (the first child of *book*), is annotated with "1.1/2" with tag type 2, which indicates *title* tag. Similarly, the *ISBN* element in first record may contain "1.2/3", followed by the *publisher* element in first record may contain "1.4/6". The rest of the nodes will be annotated with the same procedure. Figure 5 shows the XML document after the addressing process. During the addressing process, a schematic table will be built as the reference to each tag type as shown in Table 1. The schematic table is instant built upon completion of the full transverse and it is not predefined before the transversal.

Table 1.  Tag schematic table

| Tag type | Tag |
|----------|-----|
| 0 | <books> |
| 1 | <book> |
| 2 | <title> |
| 3 | <ISBN> |
| 4 | <authors> |
| 5 | <author> |
| 6 | <publisher> |
| 7 | <year> |
| 8 | <category> |
| 9 | <price> |
| 10 | <TableOfContent> |
| 11 | <Chapter> |
| 12 | <Number> |
| 13 | <Topic> |

## 3.3. Fragmentation process

In fragmentation process, a fragmentation model (horizontal, vertical, or hybrid) and a fragmentation method (structural and size, query and cost, predicate or etc.) are needed as the criteria to fragment the document. To illustrate the fragmentation process, assume we choose horizontal model and fragmentation method by size as the criteria to fragment the XML document.

Assume an XML document with 5000 book record as XML structure illustrated in Figure 3. In the addressing process, the last book record will contain a value of "5000/1" in book element *address* attribute. From this attribute we are able to get the maximum number of records in the XML document at the same level. If we would like to split these records horizontally into fragments and store in to two nodes, Then, the book elements contain value "1/1" to "2500/1" will be stored in the first node and book elements contain value "2501/1" to "5000/1" will be stored in second node.

Using the same example, if we would like to fragment with vertical fragmentation model and a fragmentation method by structure, we need to locate which child element we would like to use to split vertically. To illustrate this fragmentation, we would like to use /books/book/TableOfContent as the parameter. During the fragmentation process, element with *address* attribute that match d.d/8 where d is any positive integer number and 8 is the tag type of TableOfContent as shown in the schematic table will be fragmented out from each book record and store in node 2. The remaining data will be remained in current document or transfer to node 1.

If we would like to fragment with horizontal fragmentation model using fragmentation method of simple selection predicate by particular publisher such as "McGraw Hill", the value of the *address* attribute that match d.d/8 and the value in the element must match "McGraw Hill" in order to meet the requirement, the book record will be fragmented out from the document. Moreover, it may also allow multiple matches such as matching d.d.d/5 in author tag with the author matching the criteria. This matching will be executed faster as compare to matching a lengthy element name to locate the matching value. The matching process can be simplified by using regular expression. For instance, d.d.d/5 can be written in regular expression pattern as ^\d+\.\d+\.\d+\/5$. Selection will be perform whenever value of *address* attribute that can match this pattern.

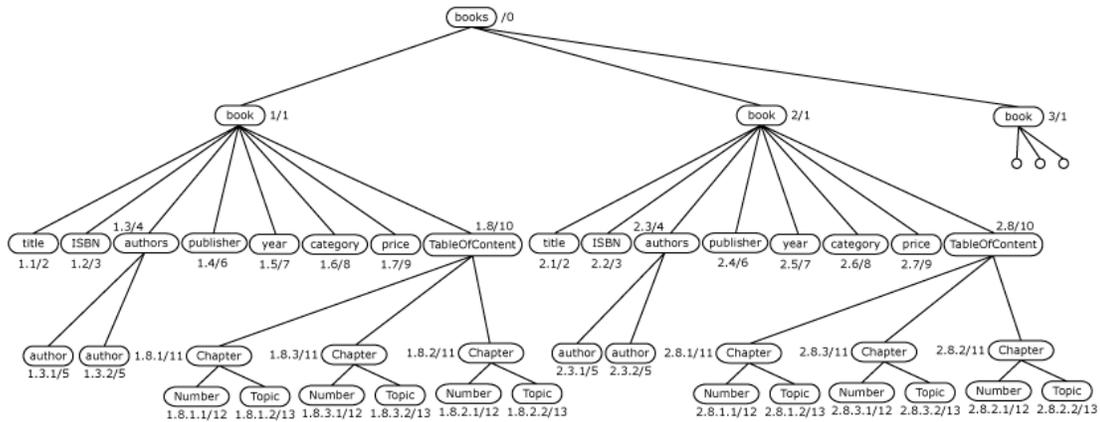

Figure 5: XML document annotated with prefix-based labelling

## 4. DISCUSSION

Structure and size fragmentation method will fragment document according to the defined structure and size of XML document. The advantage of this method will distribute the content evenly across the distributed platform. However, it does not mean effectiveness in query processing response time. Skewed query processing problem is a common problem in this fragmentation method if the query processing concentrating only on particular site or distributed nodes.

The advantage of query and cost method is the most efficient method but the fragmentation cost is higher than other two methods.

Simple selection predicate is the most fundamental fragmentation method. It works fine in fragmented large XML document into smaller pieces to reduce search time and processing power on large XML document. However, it does not work efficiently compare to the query based methods.

A fragmentation design on XML document can obviously reduce the transverse time and search time on locating particular elements in the document. Assume an XML document of records size of 100,000 is split horizontally using simple selection predicate on price into 5 nodes, the query process on particular price may only need to be performed on one node that meet the criteria range which consequently restricted to only 20,000 records.

## 5. CONCLUSIONS

XML has been widely used as an enabler for data exchange and data transfer over the Web in various applications such as electronic publishing, business transaction and application development. In many scenarios, fragments of such a repository are distributed over the Web. As such, effective way of distribution design and query models is crucial. Distribution design for databases usually addresses the problems of data acquisition, data fragmentation, allocation and replication. The focus on this paper is on data fragmentation.

In this paper, we have reviewed several techniques for distributing XML. We have reviewed the techniques in terms of fragmentation models and fragmentation operations. In addition, we have outlined our proposed fragmentation model based on prefix-labeling to annotate the nodes (subtree) of the fragment.

## ACKNOWLEDGEMENTS

This work is supported by funding of Fundamental Research Grant Scheme, from the Ministry of Higher Learning Education (MOHE).

## Authors

Kok-Leong Koong received his Bachelor in Computer Science and Master in Business Administration in University of Central Oklahoma, U.S.A. in 1995. He is currently lecturer of Department of Information Sciences and Computing Studies in New Era University College. His major area re searches are XML Databases, E-commerce, web application and computer network.

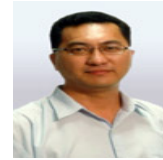

Associate Professor Dr. Su-Cheng Haw's research interests are in XML Databases and instance storage, Query processing and optimization, Data Modeling and Design, Data Management, Data Semantic, Constraints & Dependencies, Data Warehouse, E-Commerce and Web services

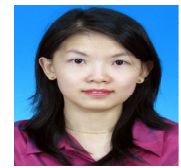

Dr. Lay-Ki Soon received her Ph.D in Engineering (Web Engineering) from Soongsil University Korea in 2009. She is currently a Senior Lecturer in Faculty of Computing and Informatics, Multimedia University. Her research interests relate to Web science, which include Web crawling, Web data mining and social network analysis. She is involved in numerous research projects funded by Malaysian government and also Japan International Cooperation Agency (JICA).

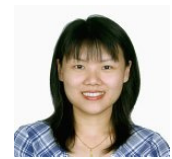

Samini Subramaniam received her Master in Information Technology from Multimedia University, Malaysia in 2011. She is currently pursuing her PhD in the area of distributed XML query processing. Her research interests are in Relational Database Management System (RDBMS), centralized and distributed XML query processing and query optimization.

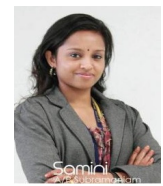